\documentclass[]{aa}

\usepackage{graphicx}
\usepackage{txfonts}
\usepackage{natbib}

\begin{document}

\newcommand{\um}{$\mu$m~}

\title{What can we learn about protoplanetary disks from analysis\\
 of mid-infrared carbonaceous dust emission?
\thanks{This work is based on observations made with the Spitzer Space
Telescope, which is operated by the Jet Propulsion Laboratory, California Institute of Technology 
under a contract with NASA.
Based on observations with ISO, an ESA project with instruments funded by ESA Member States (especially the PI countries: France, 
Germany, the Netherlands and the United Kingdom) and with the participation of ISAS and NASA.}}

\author{O. Bern\'e\inst{1,2}
	\and 
	C. Joblin\inst{1,2}
	\and
	A. Fuente\inst{3}
	\and
	F. M\'enard\inst{4}
}
\offprints{\\ O.~Berne, \email{olivier.berne@cesr.fr}}
\institute{
Universit\'e de Toulouse ; UPS ; CESR ; 9 ave colonel Roche, F-31028 Toulouse cedex 9, France
\and
CNRS ; UMR5187 ; F-31028 Toulouse, France
\and
Observatorio Astron\'omico Nacional, Aptdo. Correos 112, 28803 Alcal\'a de Henares (Madrid), Spain
\and
Laboratoire d'Astrophysique de Grenoble, Universit\'e Joseph-Fourier et CNRS-UMR~5571, 
B.P. 53, F-38041 Grenoble Cedex 9, France
}

\date{Received ?; accepted ?}

\abstract
{The disks of gas and dust that form around young stars and can lead to planet formation contain
polycyclic aromatic hydrocarbons (PAHs) and very small grains (VSGs).}
{In this Paper we analyze the mid-infrared (mid-IR) emission of these very small dust particles in a sample of 12 protoplanetary disks.
Our goal is twofold: first we want to characterize the properties of these particles
in disks and see how they are connected to interstellar matter, and second we investigate the possibility that their emission can
be used as a probe of the physical conditions and evolution of the disk.}
{We define a basis made of three mid-IR template spectra: PAH$^0$, PAH$^+$, and VSGs that were derived from the analysis of 
reflection nebulae, and an additional PAH$^x$ spectrum that was introduced recently for analysis of
the spectra of planetary nebulae.}
{From the optimization of the fit of 12 star+disk spectra, using a linear combination of the 4 template spectra, we found that
an additional small grain component with a broad feature at 8.3 $\mu$m is needed.
We find that the fraction of VSG emission in disks decreases with increasing 
stellar temperature. VSGs appear to be destroyed by UV photons at the surface of disks, 
thus releasing free PAH molecules, which are eventually ionized as observed in photodissociation regions.
In contrast, we observe that the fraction of PAH$^x$ increases with increasing 
star temperature except in the case of B stars where they are absent. We argue that this is compatible
with the identification of PAH$^x$ as large ionized PAHs, most likely emitting in regions of the disk that are close to the star. 
Finally, we provide a UV-dependent scheme to explain the evolution of PAHs and VSGs in protoplanetary disks. 
These results allow us to put new constraints on the properties of two sources: 
IRS 48 and ``Gomez's Hamburger" which are poorly characterized.}
{Very small dust particles incorporated into protoplanetary disks are processed while exposed to the intense radiation field of the
central star. The resulting shape of the mid-IR spectrum can reveal this processing and be used as an efficient probe of the radiation field
i.e. luminosity of central star.}

\authorrunning{Bern\'e et al.}
\titlerunning{PAH and VSG emission in disks}

\maketitle

\section{Introduction}\label{int}

It is widely accepted that low-to-intermediate mass ($< 8M_{\sun}$) young stellar objects (YSOs) are surrounded
by a disk of gas and dust (see e.g. \citealt{vbk04} and references therein). There is growing evidence (see
\citealt{set08} and references therein) that these disks are the cradles of planetary formation, which is why they are 
called ``protoplanetary disks". From a spectroscopic point of view, these YSOs are characterized by a 
large infrared excess due to the emission of dust present in the disk. The near-to-mid-IR region of 
their spectrum is dominated by the emission of hot dust at thermal equilibrium and present 
in the inner regions of the disk, together with the emission of stochastically
heated macromolecules and nanograins emitting from the surface layers of the disk \citep{ack04, hab04}. This emission is characterized
by ``aromatic infrared bands", generally attributed to polycyclic aromatic hydrocarbons 
(PAHs, \citealt{leg84,all85}). 

Several studies have investigated the properties of PAHs in disks. Nearly half of the
Herbig Ae/Be (2-8 M$_{\sun}$) stars observed by \citet{mee01} and \citet{ack04} show PAH emission.  Concerning their less massive counterparts,
 T-Tauri stars ($<$ 2M$_{\sun}$), \citet{gee06} tend to show that PAH emission is more marginally detected, while
\citet{bou08} report several sources with a strong emission feature at $\sim$ 8.2 $\mu$m.
These authors, as well as \citet{slo05}, show that the mid-IR PAH features undergo significant variations from one source to the next.
However, linking these variations to the modification of local physical conditions (in particular the UV field hardness) or PAH properties 
(e.g. ionization state, size) remains challenging (see \citealt{pee02, vdd04, slo07} on this subject). Recently, \citet{boe08} have
indeed argued that such modifications were linked to the chemical evolution of PAHs in the vicinity of HAe/Be stars. 
Making this connection is essential for two reasons. First, it will cast a new light on the properties of carbonaceous
particles in the interstellar dust cycle, and second, this might help in the understanding of disks evolution leading to planet formation
such as coagulation and evaporation processes. 

Unfortunately such a detailed analysis is difficult given the small spatial scales at which these processes
occur. One way to overcome this issue is by increasing the performance of mid-IR telescopes and instruments. In this field,
impressive observational \emph{tours de force} have been performed recently  revealing the mid-IR anatomy of disks \citep{lag06,gee07a,dou07}.
Another way of proceeding is by taking as much advantage as possible of the spectral information. Recently, we have studied 
the link between the shape of the observed mid-IR spectrum and the nature of emitting particles in various photodissociation
regions (PDRs) where the spatial scale is much larger \citep{rap05, ber07}. We found that their emission spectra could
generally be decomposed into 3 fundamental spectra attributed to a population of very small grains (VSGs), 
neutral PAHs (PAH$^0$), and ionized PAHs (PAH$^+$). We were able to link the presence of each one of the emitting populations
to the local physical conditions: VSGs are present in the densest regions of the clouds and dissociated
into PAH$^0$ in the more UV irradiated regions to be eventually ionized into PAH$^+$ in the cavity around the illuminating stars.
Following these studies, we propose to analyze the mid-IR emission spectra of 12 YSOs showing strong aromatic emission,
using the set of template spectra of VSG, PAH$^0$, and PAH$^+$ derived from our results obtained on PDRs.

We first describe the observations and the selected sample of objects in Sect.~\ref{obs}. 
In Sect.~\ref{ana}, we justify the choice of template spectra and proceed to the fitting of observed spectra.
In Sect.~\ref{discussion} we discuss on the nature of the different emitting populations and their link with local physical conditions
in disks. In particular we explore the general trends in the variations of the molecular (PAH) versus very small grains emission in disks as a function of
the luminosity of the central star. Conversely, we illustrate how these trends can be used to constrain the luminosity of
sources that are ambiguous in spectral type and nature, such as IRS 48 and Gomez's Hamburger.

\section{Observations and data reduction} \label{obs}


The selected sources (Table \ref{tab1}) for this study were taken from the large number of Herbig Ae/Be and T-Tauri stars that show PAH emission.
The objects in this sample are very different from one another: some are evolved and probably forming planets (e.g. HD 141569, see \citealt{aug99, mou01, bri02}), while others
are young and still probably enshrouded in the remnants of their parent cloud (e.g CD -42$^{\circ}$ 11721 see \citealt{hen98}).
The spectral type of IRS 48 is not determined well , as discussed in \citet{gee07a} and in Sect.~\ref{IRS48}.
Furthermore, the nature of Gomez Hamburger (IRAS 1805-3211) is debatable. It was originally classified as a post-AGB star \citep{rui87},
but recent observations would instead classify it as a young star surrounded by its protoplanetary disk
seen edge-on (see \citealt{buj08} and Hubble Space Telescope press release 2002/19).
We have only considered objects with prominent features detected at 6.2, 7.7, and 11.3 $\mu$m. All the sources are free of the 9.7 $\mu$m silicate 
emission feature, in order to limit the contamination of this band to the 7.7, and 11.3 $\mu$m aromatic bands. It is important to
note that selecting disks that lack silicate emission constitutes a bias in our sample. The absence of silicate
emission in the observed spectra can be due to several mechanisms: either the silicate grains are too far from the central source
and thus not heated to temperatures that are high enough to allow emission in the mid-IR, or these grains are too large to
be heated efficiently. However such effects are (at least at the first order) not related to the nature and evolution of
carbonaceous grains in disks, so we argue that this bias is minor.

\begin{table*}[ht!]
\caption{Characteristics of the source in the considered sample}
\label{tab1}
\begin{center}
\begin{tabular}{cccccc}\\
\hline \hline
 Object  & log(T$_{stellar}$) & Sp. Type & Disk evolution & Age (Myr)& Ref\\
\hline
HD 135344              &   3.810 & F4Ve & Transition & 8 $\pm$ 4& 1,3,11\\
HD 169142              &   3.914  & A5Ve & Transition & 6$^{+6}_{-3}$ & 1,4,12\\
HD 34282               &   3.941  & A3Vne &  Thin & 5 & 1,5,13\\
HD 97048 	       &   4.000  & A0/B9Ve+sh & Thick & 4$\pm$1 & 1,6\\
HD 34700               &   3.774  & G0IVe &  Thick & 3 $^{+6}_{-3}$ & 1,7\\  
RR Tau                 &   3.927  & A4e & Transition & 3$^{+1}_{-2.6}$  & 1,8,9\\
HD 141569              &   3.979 & A0Ve & Thin & 5 & 1,5\\
BD +40$^{\circ}$ 4124  &   4.291 & B2Ve & Thick+Env. & 1$^{+2}_{-0.9}$ & 1,8,14\\
CD -42$^{\circ}$ 11721 &   4.470 & B0IVep & Thick+Env. & 0.5$^{+2}_{-0.4}$ & 1,10\\
LkH$\alpha$ 215        &   4.114 & B7  & Thick+Env. & 1 & 14\\
IRS 48	               &   ?     &  ? & Thin & 4$\pm$1 & 2\\
Gomez Hamburger        &   ?    &  ? & Thick & ? & 15,16\\
\hline
\multicolumn{6}{p{12cm}}{
1: \citet{ack04},
2: \citet{gee07a},
3: \citet{dou06},
4: \citet{ram06},
5: \citet{mer04},
6: \citet{lag06},
7: \citet{tor04},
8: \citet{nat97},
9: \citet{ros99},
10: \citet{wan07},
11: \citet{bro07},
12: \citet{gra07},
13: \citet{pie03},
14: \citet{hil95},
15: \citet{rui87},
16: \citet{buj08}}

\end{tabular}
\end{center}
\end{table*}

\begin{figure}\label{Fig1}
\includegraphics[width=\hsize]{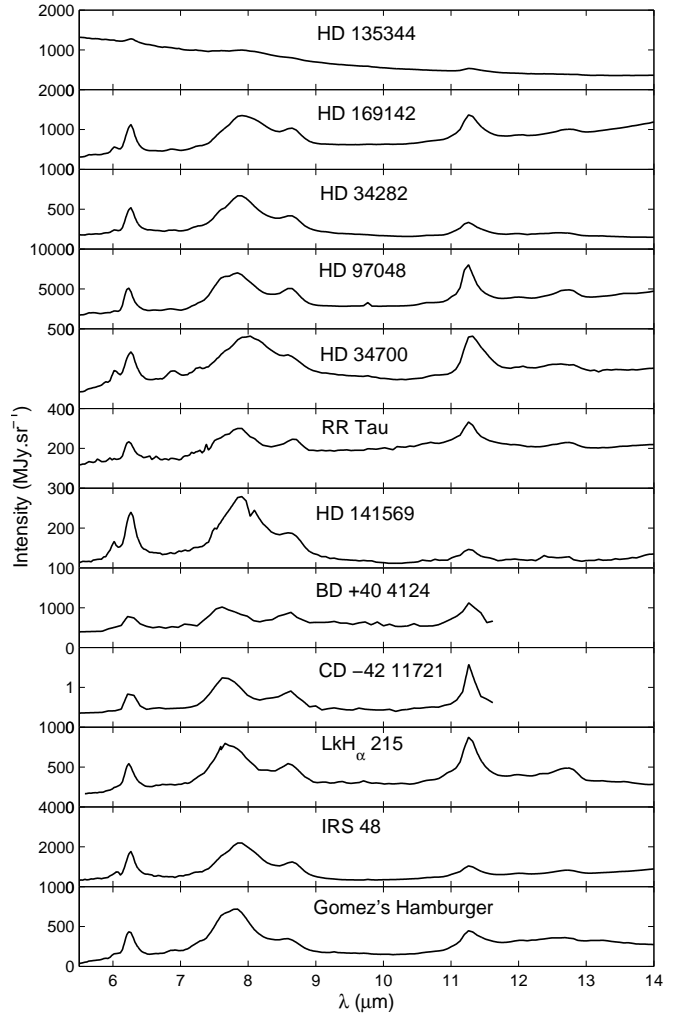}
\vspace{-0.0cm}
\caption{Mid-IR spectra of the 10 disks in our sample. All are from \emph{Spitzer}-IRS observations except BD +40$^{\circ}$ 4124 and 
CD -42$^{\circ}$ 11721, which are ISO-PHOT spectra and only cover the mid-IR range up to 11.5 $\mu$m. Intensities of the spectrum
of BD +40$^{\circ}$ 4124 were divided by 10$^3$ for clarity in the figure.}
\label{ext}
\vspace{-0.0cm}
\end{figure}

\subsection{\emph{Spitzer} Observations}

HD 135344, HD 169142, HD 34282, HD 97048, HD 34700, RR Tau, HD 141159, LkH$_{\alpha}$ 215, Gomez's Hamburger,
and IRS 48 were observed with the infrared spectrograph IRS \citep{hou04}
onboard \emph{Spitzer}, in the low resolution ($\frac{\lambda}{\Delta~\lambda}=60-127$) mode.
We have only considered the data from the short-low (SL) module covering wavelengths from 5 to 14.5 $\mu$m.
We retrieved the basic calibrated (.bcd) files processed with the S15 pipeline from the \emph{Spitzer} Science Center database.
The spectra were built using the CUBISM software \citep{smi07b} and an aperture of 4$\times$2 pixels of 1.8" to extract the spectra.

\subsection{ISO observations}

BD +40$^{\circ}$ 4124 and CD -42$^{\circ}$ 11721 were observed between 5.8 and 11.6 $\mu$m with the ISO-PHOT instrument onboard ISO, 
in spectroscopic mode ($\frac{\lambda}{\Delta~\lambda}\sim 90$). We retrieved the highly 
processed data from the ISO archive and used it directly.

\subsection{Characteristics of aromatic emission from disks}

The spectrum of each one of the sources from the sample is presented in Fig. 1. Each spectrum
clearly shows the presence of aromatic emission at 6.2, 7.7, and 11.3 $\mu$m. As described by
\citet{slo05}, the spectra resemble those defined as Class ``B" by \citet{pee02}, 
having their ``7.7" $\mu$m feature shifted towards 7.9 $\mu$m. The coolest objects, HD 135344,
HD 162149, and HD 34700, have the broadest``7.7" $\mu$m feature, and it is shifted towards 8.3 $\mu$m.

\section{Analysis of the 6-14 \um spectrum of disks} \label{ana}

\subsection{Fitting the mid-IR spectra}

Recently, \citet{job08} have applied a fitting procedure to a sample of mid-IR
spectra of planetary nebulae (PNe). The analysis employs a basis of 6 mid-IR (6-14 $\mu$m) 
template spectra to fit the continuum subtracted spectra of evolved stars: 3 PDR 
components (VSG, PAH$^0$, and PAH$^+$), very large ionized PAHs (PAH$^x$) and two
broad features centered at 8.2 and 12.3 $\mu$m.
{ The PDR components were extracted using blind signal separation methods as
described in \citet{ber07} and \citet{rap05}. The templates were then generated by
fitting these components with Lorentzians after subtraction of the rising continuum
for VSGs. They include all major features except
a plateau at $\lambda \geq 12 \mu$m for VSGs, which was difficult to reproduce due to contamination
by the H$_2$ line at 12.3 $\mu$m and underlying continuum. The PAH$^x$ spectrum was built
empirically to account for the mid-IR emission observed in PNe and H{\sc II} regions
  but was inspired by quantum chemistry calculations \citep{job08}.  As discussed in this
paper, there is some uncertainty about the intensity of the 11.3 $\mu$m band associated with
the PAH$^x$ template spectrum. The set of template spectra used in \citet{job08}
enables  the mid-IR emission arising from PNe and H{\sc II} regions
to be successfully interpreted both in the Galaxy and Small and Large Magellanic Clouds.
We therefore apply it here to the star+disk systems.}

\begin{figure}
\includegraphics[width=\hsize]{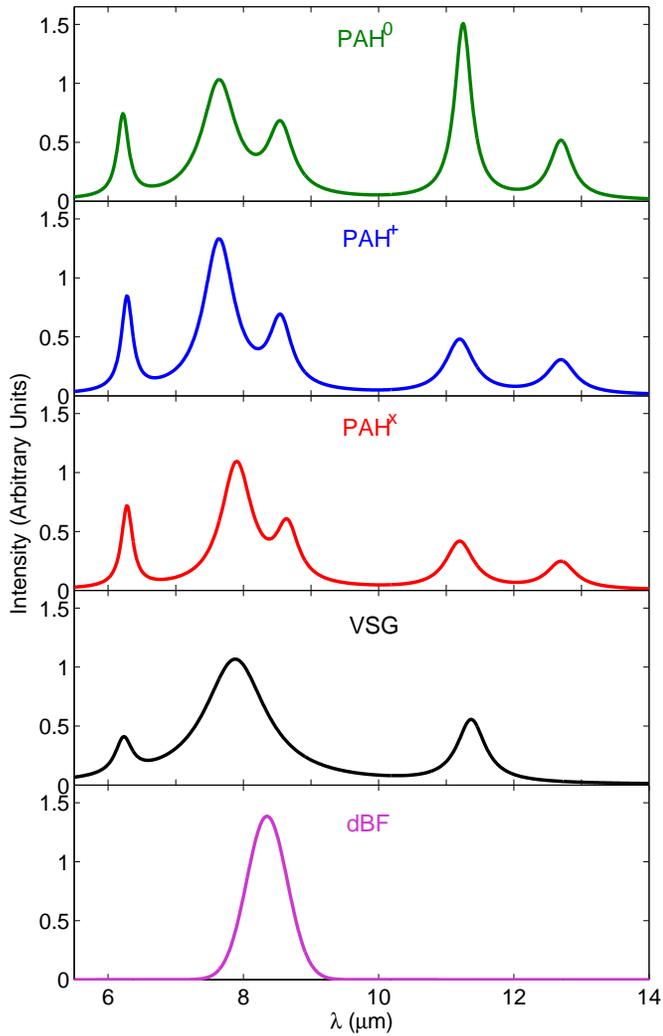}
\vspace{-0.0cm}
\caption{The five mid-IR template spectra used to fit the observed 
emission of protoplanetary disks. The PAH$^0$, PAH$^+$, and VSG spectra
are PDR components. The PAH$^x$ component (very large ionized PAHs) is 
from \citet{job08} and the 8.3 $\mu$m dBF was introduced in this paper.}
\label{basis}
\vspace{-0.0cm}
\end{figure}

In our first attempt to fit the observations, we figured out
a recurrent problem with the fits obtained for the coolest stars: the 7.7
$\mu$m massif could not be reproduced efficiently as shown in Fig. 3 for HD 135344. 
It is clear from this figure that a broad feature at $\sim$ 8.3 $\mu$m is needed to account for the observed emission.
What type of grains can be responsible for this emission? \citet{slo07} suggest that this shifted band corresponds
to the class C population defined by \citet{pee02} and is due to aliphatic bonds in emitting mixtures.
The fitting strategy we have used includes a template spectrum defined by a broad feature 
at 8.2 $\mu$m (called ``BF" in \citealt{job08}), based on a class C spectrum and prominent in post-AGB object.
The carriers of this feature are observed to be easily destroyed by UV photons,
therefore their detection in protoplanetary disks would imply that they are reformed in the parental molecular cloud
or in the densest environments of the disk itself. Since the chemical conditions differ from those prevailing 
in evolved stars there is no reason why these species should be identical in both environments, and this explains
why the 8.2 $\mu$m BF is not able to reproduce the emission of HD 135344. We therefore define a new
BF template spectrum and refer to it as dBF (for disk broad feature) to avoid confusion with the band seen in 
evolved stars. The dBF template spectrum is based on the observed spectra of T-Tauri (Fig. 14 in \citealt{bou08}).
The new fit of HD 135344 is presented in Fig.~\ref{fits}.
Following \citet{job08} we also used 2 versions of the PAH$^x$ spectrum. Results do not vary 
significantly whether the 11.3 and 12.7 $\mu$m features are present in the template spectrum or not. This implies that the shape of the 
PAH$^x$ spectrum is not well-constrained in the 11.3/12.7 $\mu$m range. It is not excluded that the carriers of the dBF also have some features in
the 10-14 $\mu$m range, but these are expected to be weak. In the following, we concentrate on the results including the 11.3
and 12.7 $\mu$m feature in the PAH$^x$ spectrum, and no additional feature in the 8.3 $\mu$m dBF spectrum. 
As discussed in \citet{job08}, the main effect of using a PAH$^x$ spectrum without 11.3 and 12.7 $\mu$m feature will be to 
slightly decrease the fraction of PAH$^0$ in the results of the fits.
Finally, we note that the 12.3 $\mu$m BF used in \citet{job08} is useless in the fits of protoplanetary disks, so
we do not use it here. The final basis of 5 template spectra we have used for the fitting procedure
is presented in Fig.~\ref{basis}.

Considering the geometry and density of disks, IR radiative transfer might be a concern. We have therefore re-run our fit
using simple extinction parameter based on the extinction curve of \citet{wei01}, for a total-to-selective extinction ratio R$_V$=5.5.
Though in some cases the fit is slightly better, the estimated fraction of each population is nearly unchanged. Thus , for the sake of simplicity,
we present here results that do not include extinction.

\begin{figure}\label{bad_fit}
\includegraphics[width=\hsize]{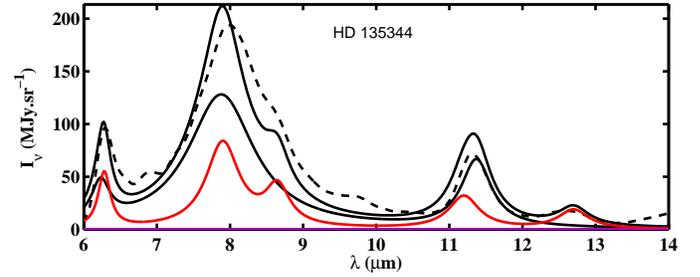}
\vspace{-0.0cm}
\caption{Preliminary fit of HD 135344, using only PDR components together with PAH$^x$. Observed spectrum is the
upper dashed line and fit is the continuous line. The contribution of VSG, PAH$^x$, and PAH$^0$ is represented below
in black, red, and green, respectively. It appears clearly that the use of these components cannot reproduce
the observed emission and that a new component with a feature at 8.3 $\mu$m is needed (see dBF in Fig.~\ref{basis}).
The new fit is presented in Fig.~\ref{fits}}
\label{ext}
\vspace{-0.0cm}
\end{figure}

\subsection{Results} \label{res}

\begin{figure*}[ht!]
\begin{center}

\includegraphics[width=8cm]{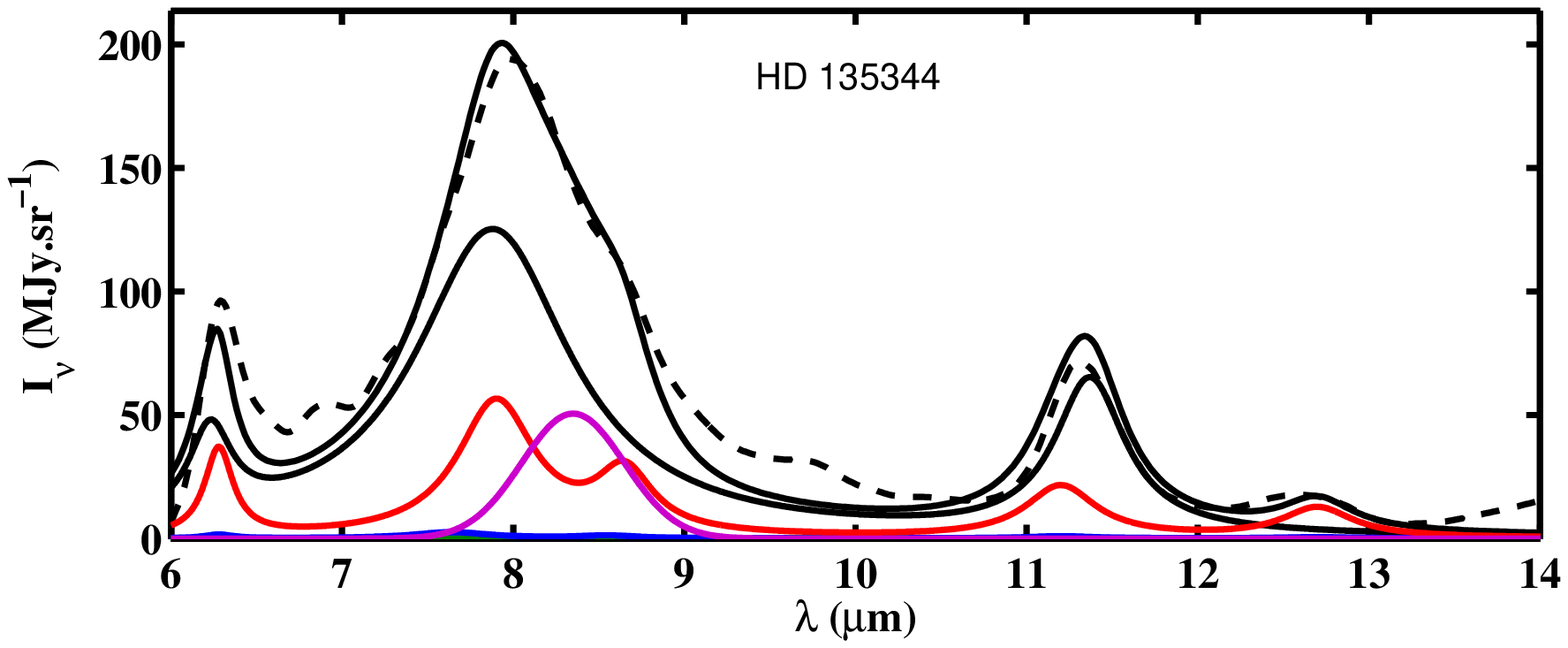}
\includegraphics[width=8cm]{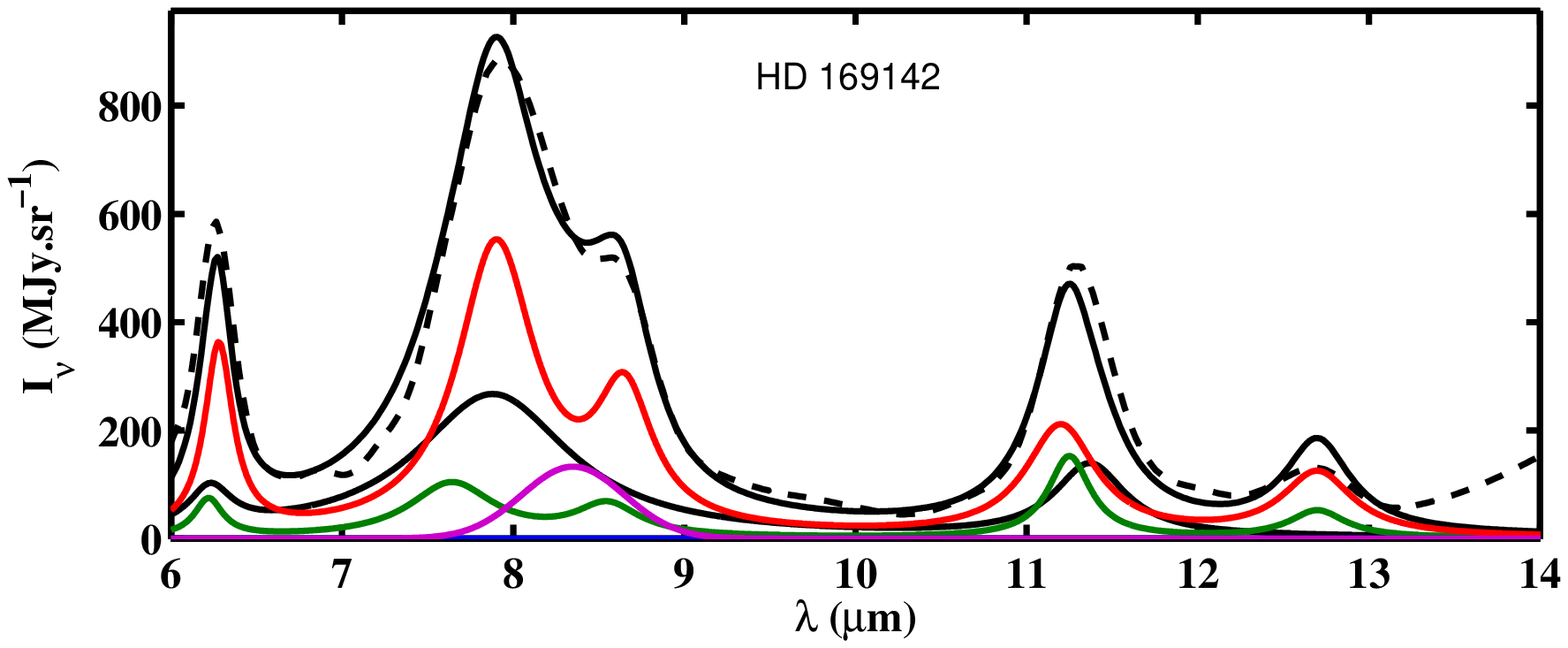}
\includegraphics[width=8cm]{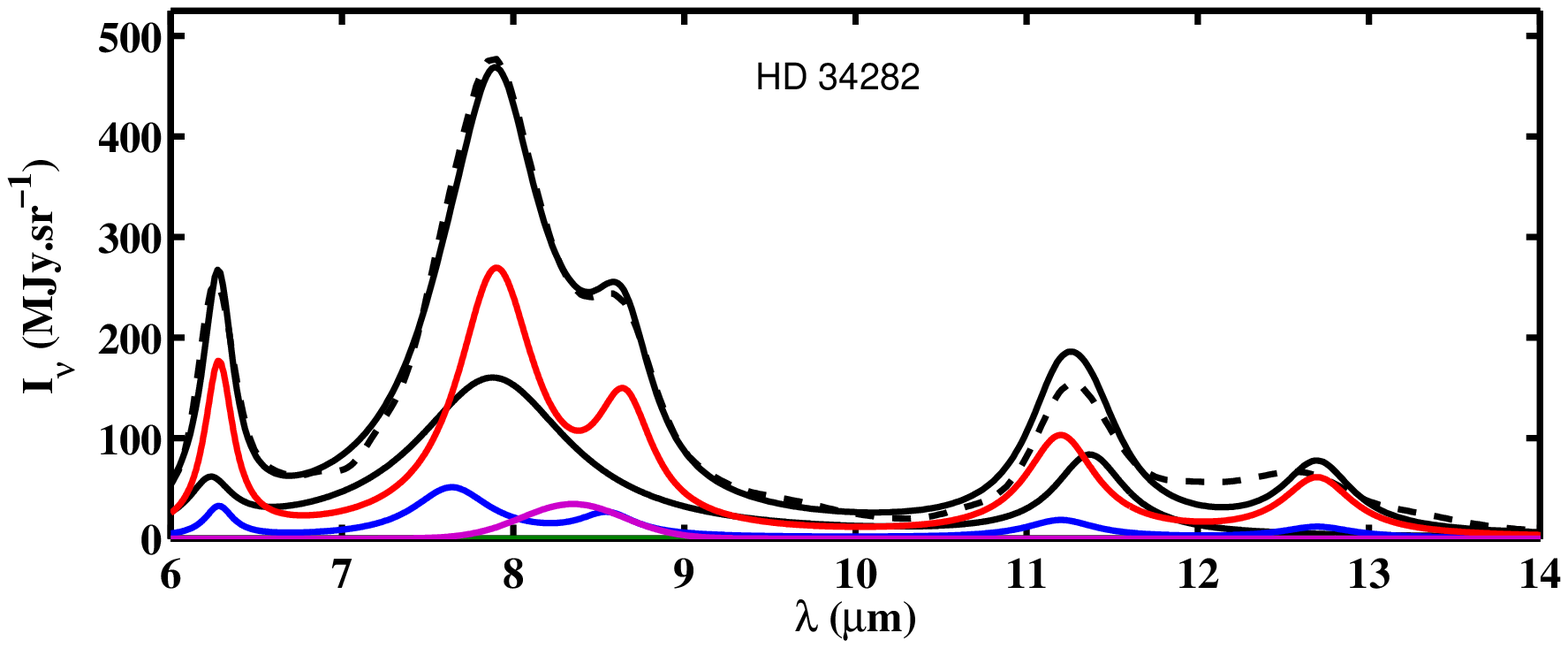}
\includegraphics[width=8cm]{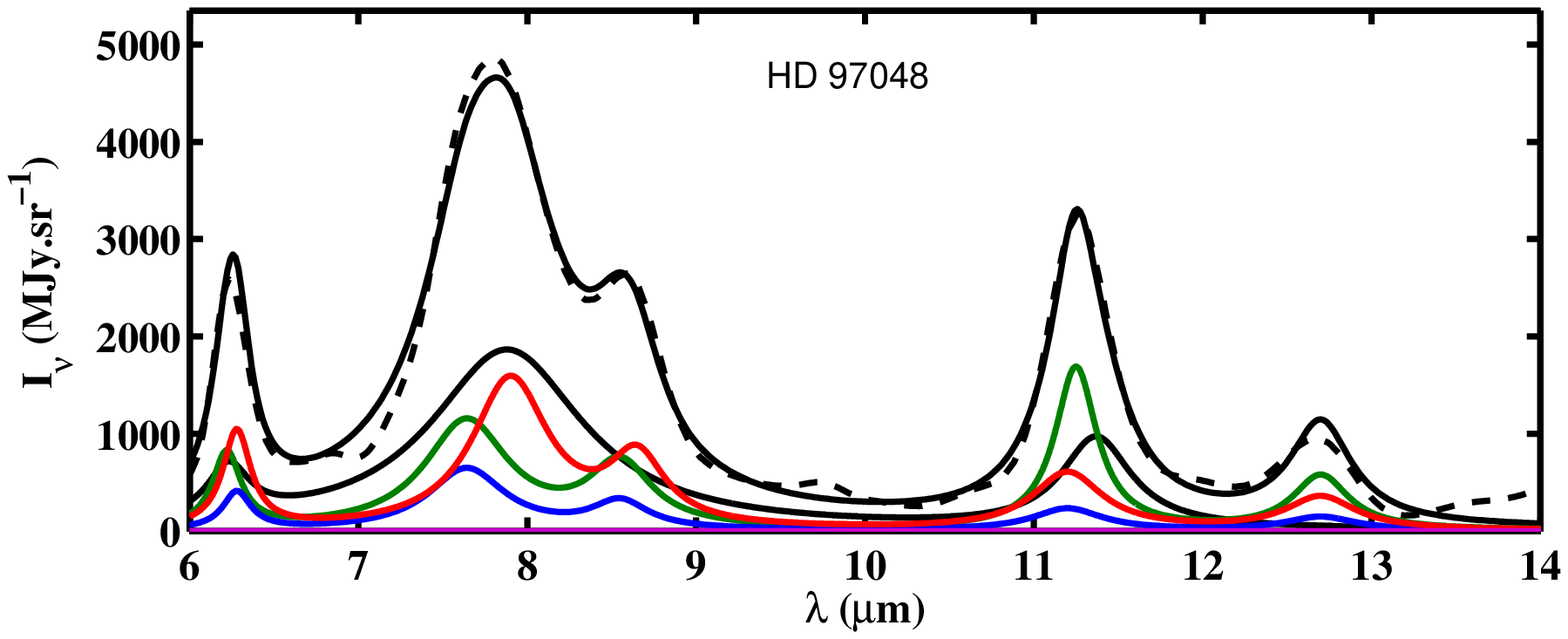}
\includegraphics[width=8cm]{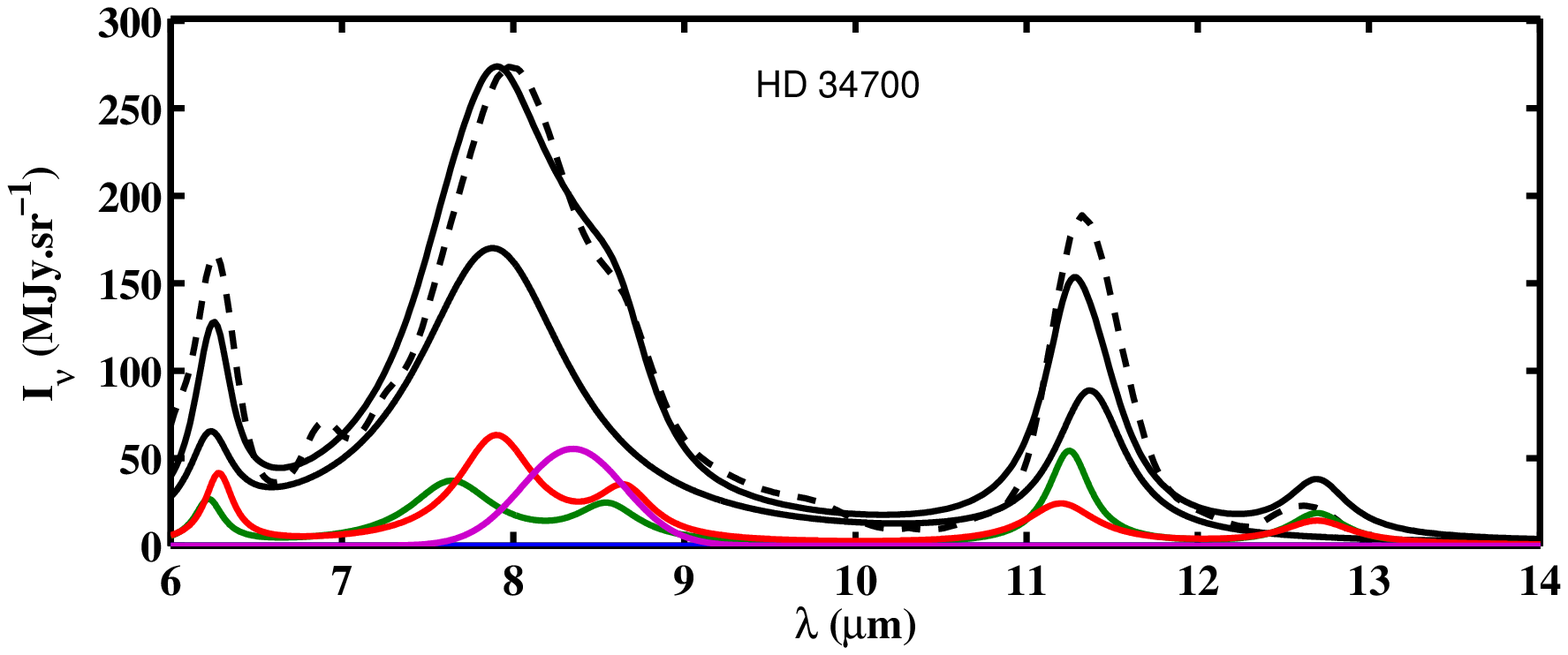}
\includegraphics[width=8cm]{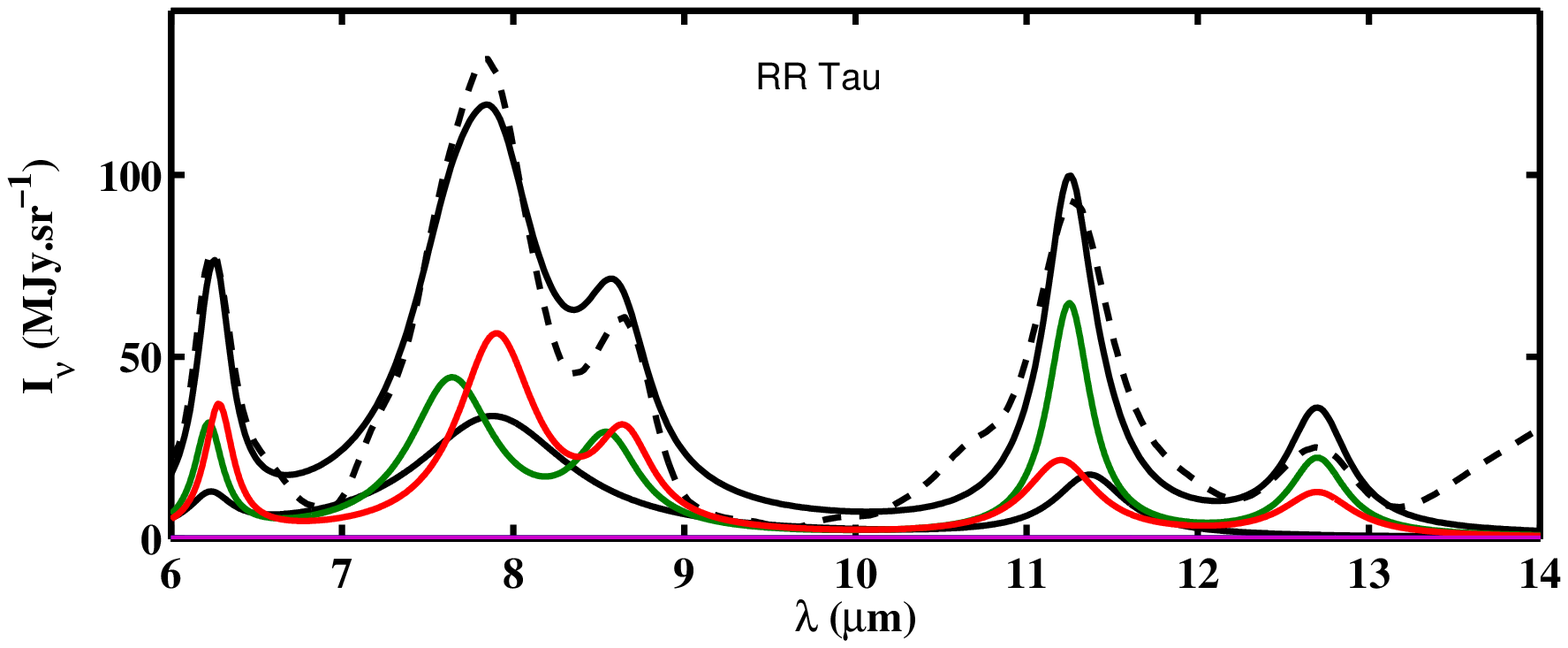}
\includegraphics[width=8cm]{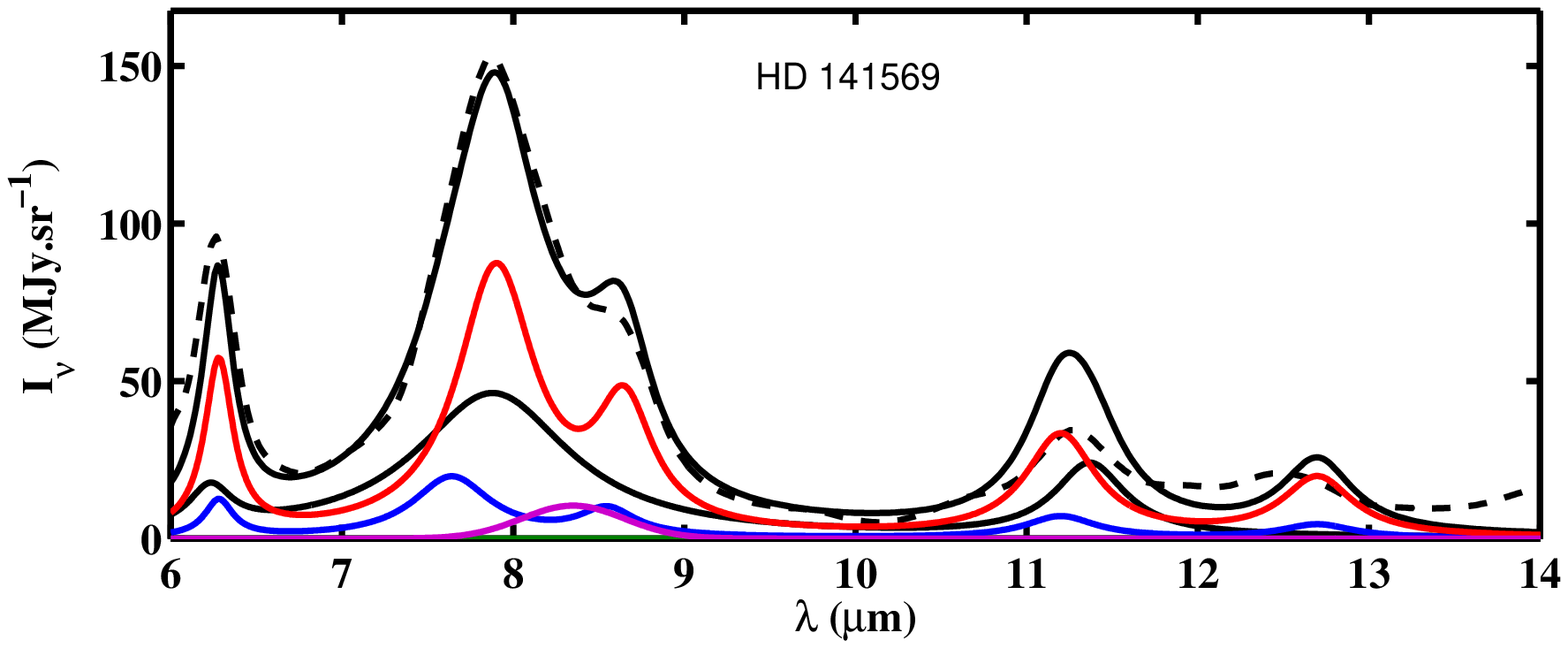}
\includegraphics[width=8cm]{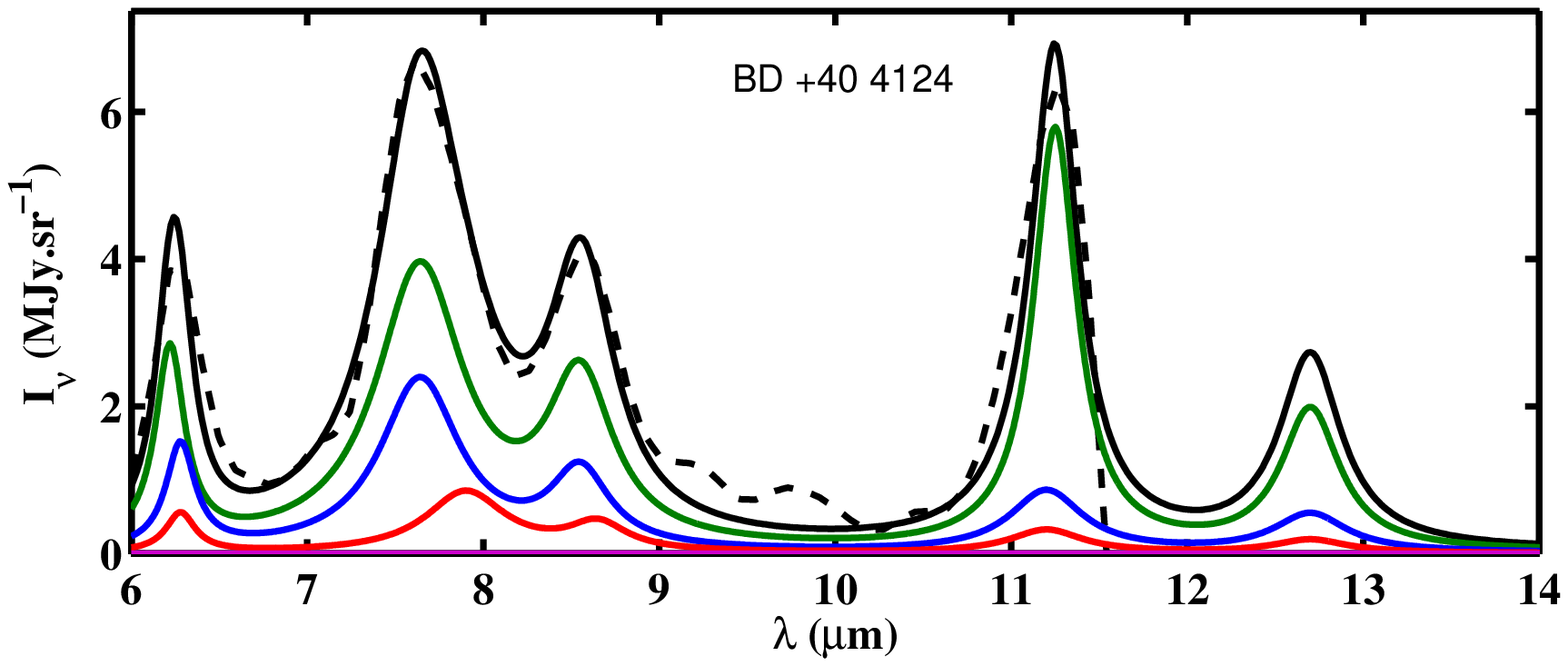}
\includegraphics[width=8cm]{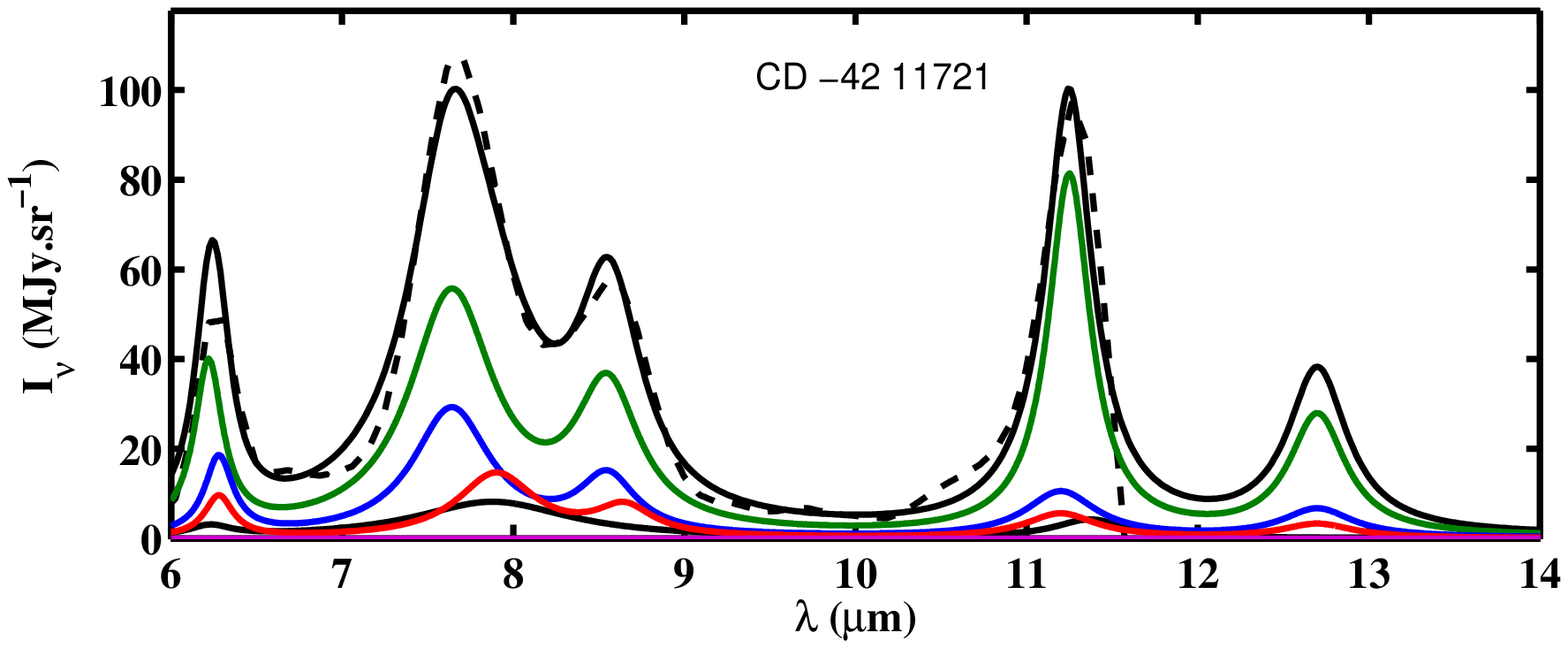}
\includegraphics[width=8cm]{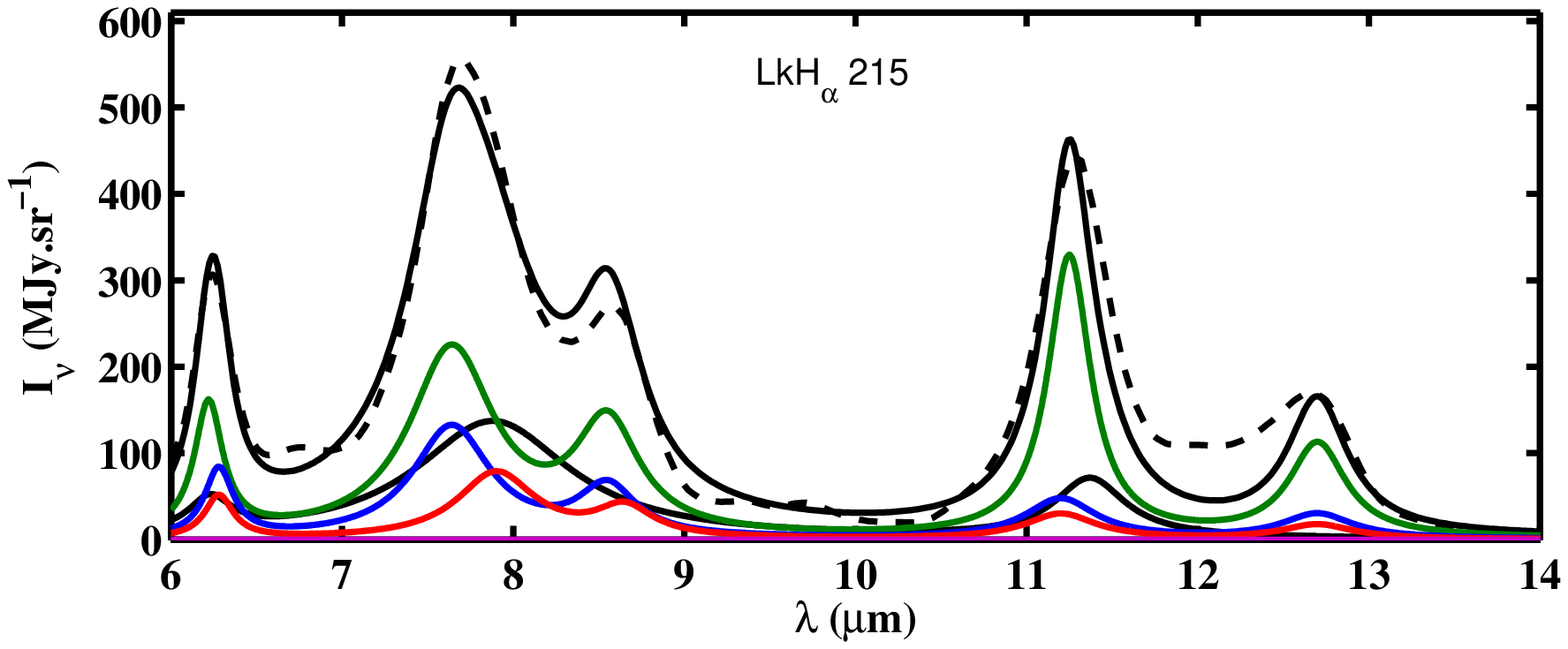}
\includegraphics[width=8cm]{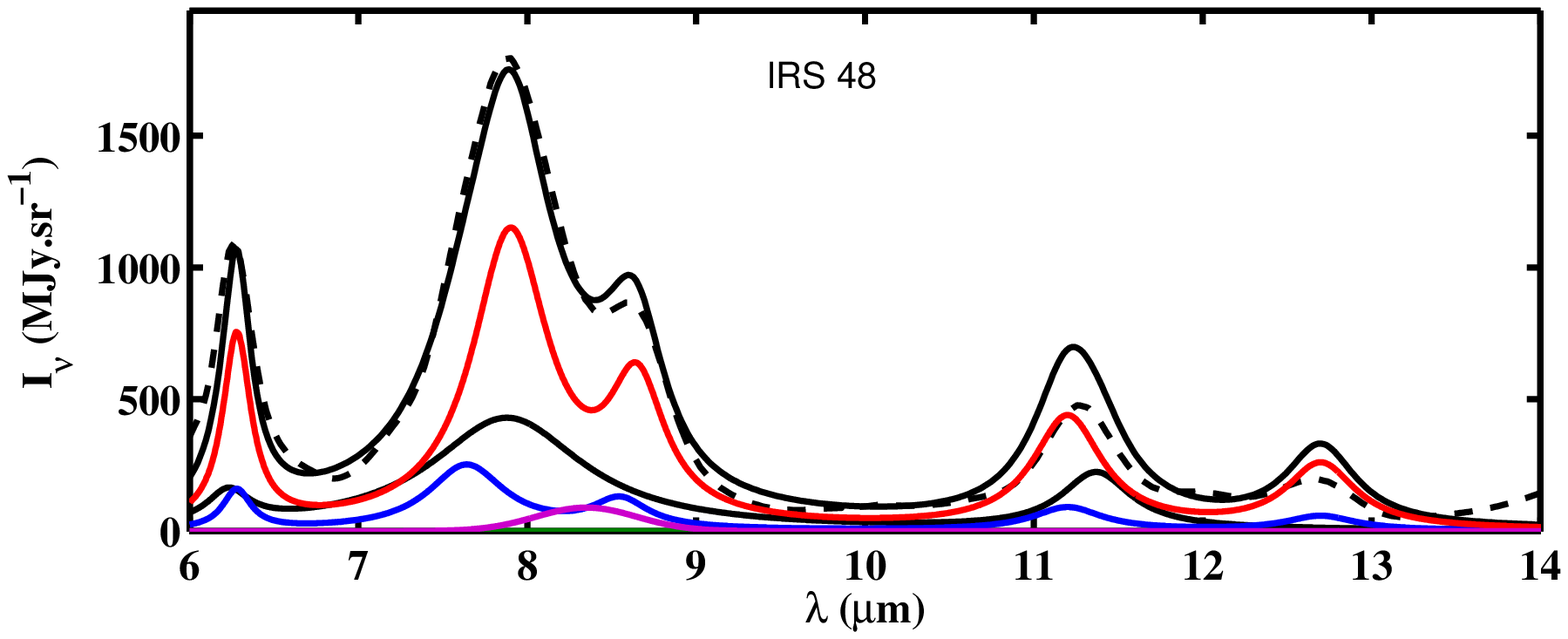}
\includegraphics[width=8cm]{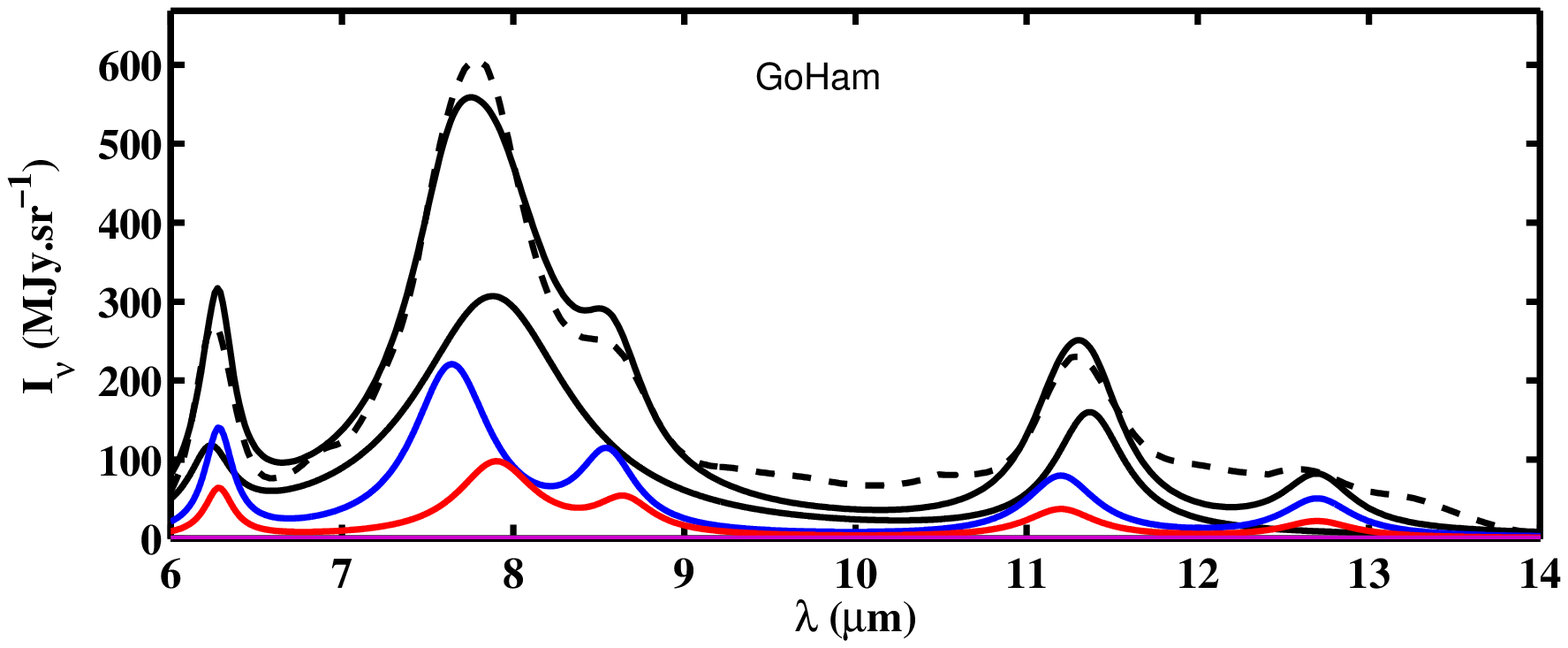}

\caption{Fits of the 12 spectra of the star+disk systems observed with \emph{Spitzer}-IRS and ISO using our set of
5 template spectra. The observed spectrum is the dashed black line, the fit is
the continuous black line. The contribution of each component is shown below in colors (using the same
color code as in Fig. 2.) }
\label{fits}
\end{center}
\end{figure*}

The results of the fits are presented in Fig.~\ref{fits}.
For each object we are able to reproduce the observed spectrum quite efficiently. 
Table \ref{tab2} gives the fractions of mid-IR integrated fluxes for each component present in the studied source. 
Because mid-IR emitters are heated and processed by the UV-visible photons emerging
from the central star, we propose to compare the fractions of each emitting population 
to the theoretical total and UV luminosities of the central source given its spectral type (see Table 1). 
We used the spectra from the 1993 Kurucz stellar atmosphere atlas (http://www.stsci.edu/hst/observatory/cdbs/k93models.html
based on \citealt{kur79}). We used a gravity of 4 for all sources and a stellar radius 
from \citet{all82}. The total luminosity is defined as $4\pi R_{*}\int F_{\lambda} d\lambda$
and the UV luminosity as $4\pi R_{*}\int_{91.2 nm}^{240 nm} F_{\lambda} d\lambda$ (according to \citealt{hab68}), where $R_*$ is the
stellar radius and $F_{\lambda}$ the flux from the Kurucz catalog.
In Fig.~\ref{VSG} we plot the fraction of VSG emission as a function
of UV luminosity, showing a clear decrease of this emission while the mass
of the star increases. In contrast, the fraction of PAH emission increases with the luminosity 
(Fig.~\ref{PAH}-\ref{PAHxs}). This is due to the increasing PAH$^x$ emission, whereas the PDR-type 
PAH emission is found to increase up to spectral type A4 and then to decrease when going towards stars of
earlier spectral types (Fig.~\ref{PAHs}). The same trends are observed either using total or UV luminosity.
We also observe that there are no dBF in the spectra of disk around stars of spectral type earlier than A0 (i.e. UV luminosity $> \sim 10 ^{28}$ W).

\begin{table}[h!]
\caption{Factions of each population of the carbonaceous dust emission in the spectrum
of the 12 sources of the selected sample}
\label{tab2}
\begin{center}
\begin{tabular}{lcccccc}
\hline \hline
\noalign{\smallskip}
& \multicolumn{5}{c}{fraction of the emission ($\%$)}\\
\cline{2-6}
 \noalign{\smallskip}
 Object      & VSG & PAH$^0$ & PAH$^+$ & PAH$^x$ & dBF \\
\hline

HD 135344    &   66        & 0   &   1       &   24     &   9  \\
HD 169142    &   38        & 25  &   11      &   26     &   0  \\
HD 34282     &   30        & 13  &   0       &   51     &   6 \\
HD 97048     &   37        & 0   &   10      &   50     &   3  \\
HD 34700     &   61        & 14  &   0       &   18     &   7  \\
RR Tau       &   27        & 37  &   0       &   36     &   0  \\
HD 141569    &   34        & 0   &   12      &   51     &   3  \\
BD+40 4124   &   0         & 61  &   29      &   10     &   0  \\
CD-42 11721  &   8         & 57  &   23      &   12     &   0 \\
IRS 48       &   27        & 0   &   13      &   58     &   2  \\
GoHam        &   54        & 0   &   32      &   14     &   0  \\
LkH$\alpha$215     &   25        & 43   &   20       &   12    &    0  \\

\hline

\end{tabular}
\end{center}
\end{table}

\section{Probing disks with mid-IR emission ?}\label{discussion}

\subsection{The origin of the 8.3 $\mu$m broad feature}

The 8.3 $\mu$m dBF component is found primarily in the sources with the coolest stars, which suggests 
a fragile material as in post-AGB stars, possibly of aliphatic nature. 
\citet{slo07} suggest that the grains responsible for the 8.3 $\mu$m feature in cool protoplanetary disks 
are similar to the grains responsible for the``class C" spectrum of \citet{pee02} and observed in post-AGB stars. 
However, we find that the dBF position and width are not compatible with the class C spectrum, the dBF
being broader and peaking at 8.3 rather than 8.2 $\mu$m, as also noticed by \citet{bou08}.

\subsection{Processing of VSGs}\label{vsgevol}

\begin{figure}[h!]
\begin{center}
\includegraphics[width=\hsize]{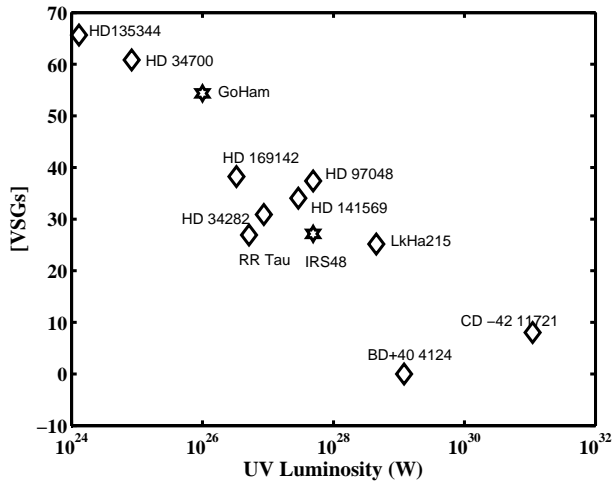}
\caption{Fraction of VSG emission in the selected sources, as a function of
the UV luminosity of the central star. Diamonds are for sources that have a well-defined spectral type. The spectral types
of GoHam and IRS 48 (shown with star symbols in this plot) are inferred in order to agree with the
found correlation. The correlation shows that VSGs are less abundant around hot stars,
likely due to their rapid destruction in these harsh environments. }
\label{VSG}
\end{center}
\end{figure}

The fraction of VSG emission clearly decreases with increasing UV luminosity (see Fig.~\ref{VSG}), 
while emission from free PAH molecules becomes more and more prominant. This is in line with observations of PDRs \citep{ces00, rap05} 
where it appears that VSGs are dissociated into free PAH molecules. It is expected that such a
process is also taking place on the surface of the disk. Still, one has to consider the competition between 
evaporation and coagulation \citep{rap06}, but evaporation is likely to dominate
when the central star becomes hotter and therefore produces more and more UV photons.
\citet{dul07} show that in more evolved disks the ratio of PAH-emission over far-infrared emission 
should be much higher than what is observed, because of the lower optical
thickness of the disk in the wavelength range at which PAHs absorb.
Thus they propose that coagulation of PAHs could be the reason for this lack of emission.
Rapacioli et al. (2005, 2006) suggest that VSGs are indeed coagulated PAHs. In this scenario, an increase in the
abundance of coagulated PAHs would imply an increase in VSG emission in more evolved disks
which is not observed. Instead, one could claim that this lack of PAH emission is due to the geometrical
evolution of the disks. Indeed, an older and flatter disk will receive less UV photons per unit of surface,
and thus the heating of PAHs (if still present) might become less efficient leading to a less intense
mid-IR signal.

\begin{figure}[h!]
\begin{center}
\includegraphics[width=\hsize]{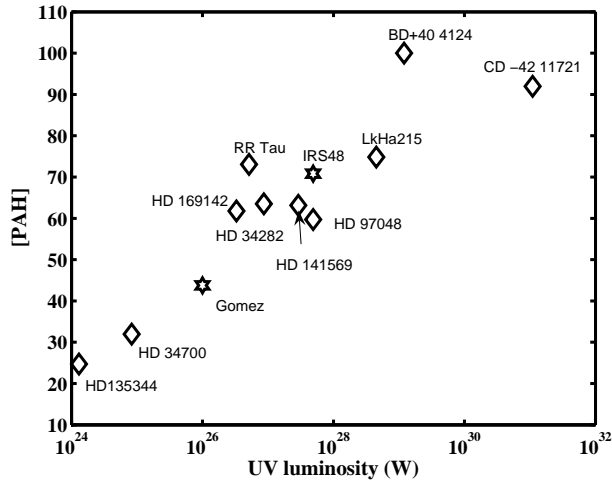}
\caption{Fraction of PAH (PAH$^{0}$+PAH$^{+}$+PAH$^{x}$) emission in the selected sources, as a function of
UV luminosity. Diamonds are for sources that have a well-defined spectral type. The spectral types
of GoHam and IRS 48 (shown with star symbols on this plot) are inferred  to agree with the
found correlation.}
\label{PAH}
\end{center}
\end{figure}

\subsection{Evolution of PAH populations}\label{pahevol}

Figure~\ref{PAHs} presents the fraction of PDR-type PAHs as a function of the stellar UV luminosity.
There is a clear rise in the fraction of emission from sources of spectral type G0 to A4, and then
this fraction decreases again with increasing luminosity. The rise can be interpreted as due to the
high production rate of PAHs from VSGs (see previous section) in the disks around stars of G0 to A4 spectral types.
For sources that are hotter, these molecules start being destroyed by the  strong UV-visible radiation field,
 which explains the decrease in the observed PDR-type PAH emission fraction. In contrast, the fraction of PAH$^x$
emission increases (Fig.~\ref{PAH}). 
\citet{job08} conclude that this is the  likely result of the destruction of the smallest PAHs combined with
efficient heating of the PAH$^x$ population. This is consistent with PAH$^x$ being very large ionized PAHs that can better survive
in extreme irradiation conditions.

The charge state of PAHs in disks is determined by the competition between ionization rate versus electron recombination for cations/neutrals
and photodetachment versus attachment rate for anions/neutrals. In disks, the gas density can be as high as 
10$^{7-8}$ cm$^{-3}$, and the electron attachment to PAHs can be more efficient than 
re-ionization by UV radiation as shown by \citet{li03}. This would imply that PAHs can be found as anions (PAH$^-$) that could
contribute strongly to the mid-IR emission spectrum of disks. In this case it is tempting to identify PAH$^x$
as very large PAH$^-$. However \citet{vis07} show that in flared optically 
thick disk (e.g. HD 97048), and even though anionic PAHs could be abundant, they are likely to be situated in deep regions of the disk where 
they receive too little UV light from the star to be excited so they dominate the mid-IR emission. On the other hand, they suggest that
the emission at the surface of HAe/Be disks will arise from a mixture of very large ($>$ 100 C atoms), 
positively ionized PAHs. 

\begin{figure}[h!]
\begin{center}
\includegraphics[width=\hsize]{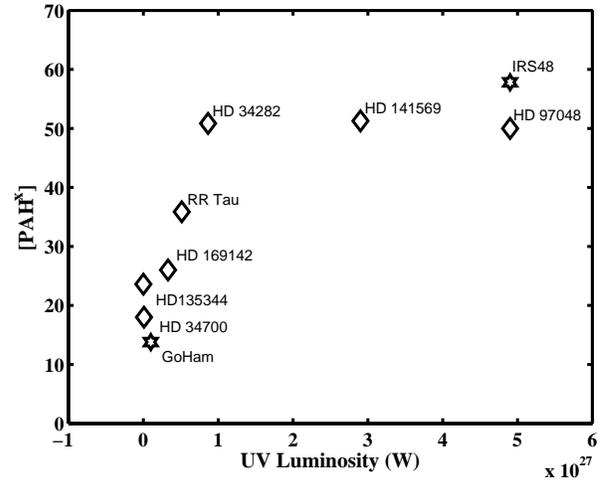}
\caption{Fraction of PAH$^x$ in the disk of T-Tauri and HAe sources of the sample.
Diamonds are for sources that have a well-defined spectral type. The spectral types
of GoHam and IRS 48 (shown with star symbols on this plot) are estimated to agree with the
found correlation.}
\label{PAHxs}
\end{center}
\end{figure}

\begin{figure}[h!]
\begin{center}
\includegraphics[width=\hsize]{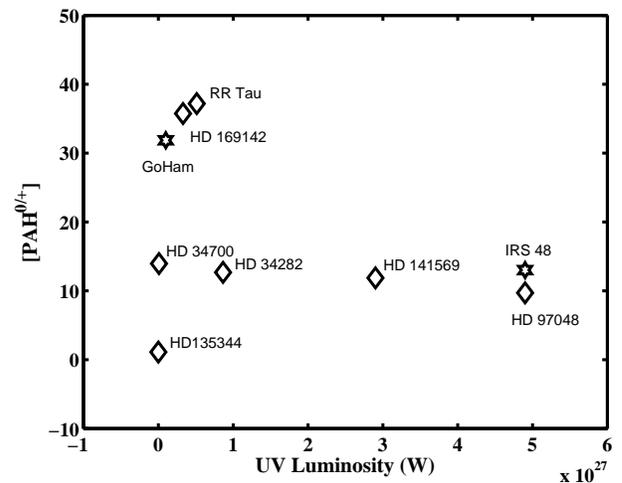}
\caption{Fraction of PDR-type PAHs (PAH$^+$+PAH$^0$) in the disk of T-Tauri and HAe sources of the sample.
Diamonds are for sources that have a well-defined spectral type. The spectral types
of GoHam and IRS 48 (shown with star symbols on this plot) are estimated to agree with the
found correlation.}
\label{PAHs}
\end{center}
\end{figure}

Figure~\ref{PAHxs} shows that the fraction of PAH$^x$ emission increases with UV luminosity 
for the studied T-Tauri and Herbig Ae stars.
For the Be star nearly no PAH$^x$ emission is detected, which we interpret as due to the absence of disk emission in 
the observed spectra (see Sect.~\ref{Be}), so it is not plotted on Fig.~\ref{PAHxs}.
Furthermore, these objects do not exhibit the dBF and their spectrum can be 
reproduced using  the PDR spectra almost exclusively. \emph{Spitzer}-IRAC images as well as millimeter 
observations \citep{hen98} suggest that these sources are not isolated, and their mid-IR spectrum was indeed previously 
classified as ``A" by \citet{vdd04}, which is consistent with typical PDRs. This suggests that we are in fact not 
observing the emission from the disk but from the PDR created in the nebulosity around the central source.
\citet{boe08} come to the same conclusions concerning the B9 stars HD 37411 and HD 36917 observed with \emph{Spitzer}, for which
they suggest that the observed emission is arising from the surroundings of the star but not a disk.
There can be three reason for this: (1) no PAH can resist in the disk because they are all photodissociated,
(2) there is no disk because it has been already photoevaporated, (3) the geometry of the disk is such 
that the PAH emission is not sufficiently activated by UV-visible photons. We discuss these scenarios in Sect.~\ref{evol}.

That some PAH$^x$ emission is present in cool stars, as in the F4 star HD 135344, implies that
the near-UV radiation field is sufficient to destroy some of the smallest PAHs and induce PAH$^x$ emission.
This means that PAH$^x$ are close to the star despite HD 135344 having a rather large inner radius 
(see e.g. \citealt{pon08}). In this case, one could consider that PAHs are in fact coupled with a gas disk 
that extends closer to the star than the thick disk \citep{pon08}.
The presence of PAH$^x$ in HD 135344 argues against the possibility that these species are positively ionized as
the F4 star is unlikely to produce enough UV to ionize the PAHs.
On the other hand, PAH cations appear to be good candidates considering that they are very abundant in emission
in the disk around A stars that produce a large amount of UV (Fig.~\ref{PAHxs}).  
As in \citet{job08} we conclude that PAH$^x$ are very large ionized PAHs tracing harsh irradiation conditions.
Unfortunately, the observational constrains provided here are not sufficient to determine if PAH$^x$ correspond to a population
of cations or anions or even a mixture of both. We emphasize nevertheless that the detection of PAH$^x$ around a young star provides
a way to assess the presence of a disk.

\subsection{The mid-IR spectrum: a probe of the irradiation
conditions of the disk}\label{evol}

\begin{figure}
\begin{center}
\includegraphics[width=\hsize]{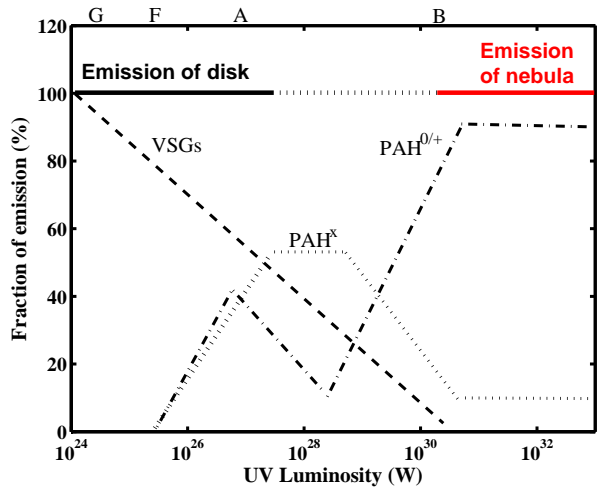}
\caption{Schematic representation of the contributions of the very small dust particle populations to the mid-IR 
spectra of T-Tauri and Herbig Ae/Be stars. While the spectrum of G, F, 
and A stars is dominated by disk emission, B star spectra are likely due to emission
arising from the remnants of the parental cloud. The total emission is due to 
the contribution of VSGs (dashed line), PDR-type PAHs (dash-dotted line), and PAH$^x$ (dotted line).}
\label{schema}
\end{center}
\end{figure}

Finally, can we trace the morphology/content of the disks using the spatially non-resolved mid-IR \emph{Spitzer} spectra?
The answer is probably no, because the evolution of very small dust particles in disks is clearly
connected to the irradiation conditions and thus to the spectral type of the central star.
This connection between the composition of the mid-IR spectrum and spectral type
is presented in a schematic way in Fig.~\ref{schema} and corresponds to the following
scenario: the dense disk is a reservoir of VSGs (such as molecular clouds in PDRs), which are constantly dissociated into free PAH
molecules. The more intense the UV field (i.e. the more massive the star is), the shortest
the lifetime of a VSG at the surface of the disk, where the emission is occurring. This is
why the fraction of VSG emission decreases with inreasing star temperature (Fig.~\ref{schema}). Similarly,
once they are released from VSGs, the lifetime of the smallest PAHs around hot stars is too short to allow their emission to be 
dominant. This is why the fraction of large PAHs (PAH$^x$) that resist destruction 
more is greater in disks around hotter stars (Fig.~\ref{schema}). More massive B stars only show 
emission of PDR-type PAHs. This might be because this emission does not arise predominantly
from a disk, either because VSGs and PAHs (even very large PAH$^x$) do not survive in disks around Be stars
or because the disk is small or absent (see next section).

\subsection{Photoevaporation of disks around Be stars}\label{Be}

We have shown that Be stars have a PDR-type emission and that the PAH$^x$ emission is not observed 
in these objects.
This supports the idea that there are no disks around these stars or that the disk emission
in the mid-IR is too weak to be disentangled from the surrounding PDR.
\citet{vis07} have shown (see their Fig. 3) that a radiation field of more than $\sim 5\cdot10^7$ in 
units of the Habing field will destroy even large PAHs of 100 carbon atoms in less than 3 Myrs. 
An estimate of the radiation field at a distance of 300 AU from a B2-0 star, taking 
only dilution effects into account  and no UV radiative transfer, gives values as high as 10$^{8-9}$. This is 
high above the limit value of $5\cdot10^7$, so it is likely that in such a harsh environment most, if 
not all, PAHs will be destroyed efficiently at a radius of 300 AU and even more efficiently at smaller radii.
A second possibility is that the disk itself is absent because it was photoevaporated. It is indeed believed that
UV can provoke the dispersion of the gas and smallest dust particles (therefore probably including PAHs and VSGs)
within a few 10$^5$ Myr (see e.g. the recent review by Alexander, 2007).
Recently, \citet{alo08} have found that the dusty disks around Herbig Be stars tend to be lighter 
and smaller than those towards cooler stars, probably due to the photoevaporation effect. The PAH emission 
in such a small disk is thus expected to be very weak in any case (whether or not PAHs are destroyed). 
Note that the effect of PAH destruction in the disks around Be stars will also directly affect 
photoevaporation of the disk. PAHs and VSGs are responsible for most of the UV-optical 
thickness of the disk \citep{dra03} so their destruction would imply a decrease in the
UV optical thickness. This effect combined 
with the higher proportion of UV photons, will tend to accelerate the photoevaporation process 
of the disk and implies that the lifetime of disks around massive stars could be shorter than
the time needed to disperse the surrounding material as suggested by \citet{alo08}.
The probably flatter shape of the disk around Herbig Be stars (because of 
the photoevaporation of the outer layers) could also contribute to the 
absence of PAH emission because of the self-shadowing effect.

\section{Putting new constraints on the nature of IRS 48 and Gomez's Hamburger}\label{probe}

\subsection{The luminosity of IRS 48}\label{IRS48}

The spectral type of IRS 48 is controversial (see discussion in \citealt{gee07a}).
Here we propose to classify it using the results of our analysis. With the observed
abundance of PAH$^x$, VSGs, and the presence of the dBF, we can estimate in which range of temperature
IRS 48 falls. The PAH$^x$ fraction in the emission of IRS 48 is 58\%, which
is compatible with a luminosity of $\sim$ $5\cdot10^{27}$ W based on Fig. 4. 
The VSG fraction is 29 \%, which also compatible with a UV luminosity of $\sim$ $5\cdot10^{27}$ W
(Fig. 5). Finally, the fraction of dBF in the spectrum of IRS 48 is 2\%, which puts it near the limit
for the presence of the dBF, consistent with a UV luminosity of $\sim$ 10$^{27-28}$ W.
Furthermore, we observe a 7.7 $\mu$m to 11.3 $\mu$m band ratio above 2. Using \citet{fla06}, this implies a
$\frac{G_0 T^{1/2}}{n_e}$ above 10$^5$.  Using a temperature T of 500 K and an electron 
density of n$_{e}$=$10^4$ cm$^{-3}$ at the surface of the disk \citep{jon07, vis07}, 
this leads to a UV field of $G_{0} \sim 60 000$ in units of the Habing
field, which is similar to what is found in the emitting regions of disks around an A0 star.
This value, as well as the derived luminosities, are compatible with the
powering source being an early A star. If the spectral type of the 
source is really around F3 or even M0 \citep{gee07a}, an excess of UV is needed and might
indicate accretion processes.

\subsection{The nature of Gomez Hamburger}

Gomez Hamburger (GoHam hereafter) is subject of debate in terms of classification. It was
originally classified as a post-AGB star \citep{rui87}, and recent observations would
instead classify it as a pre-main-sequence (PMS) A star surrounded by its disk \citep{buj08}.
We propose to investigate here the observed mid-IR spectrum of GoHam is compatible
with the PMS-star + disk hypothesis. As described above, we can use the results of
the fitted spectra to investigate the properties of the illuminating star. The found
fractions of VSGs and PAH$^x$ and dBF carriers are compatible with a total luminosity of $\sim$ 8
L$_{\sun}$ and a UV luminosity of $\sim$ 0.3 L$_{\sun}$. This indicates that the central
star should have an effective temperature of $\sim$ 7000-9000 K. This is slightly cooler than
the stellar temperature  of $\sim$ 10 000 K derived by \citet{rui87}  from near-IR photometry.
However, one should note that measuring the temperature of the star using
photometric bands is extremely difficult and unreliable given that the disk is seen edge on,
and thus that the observed light is completely scattered by the disk.
Based on our analysis, the central star in GoHam would be an A5-8 star near main 
sequence (no emission lines are present in the visible spectrum observed by \citealt{rui87}). 
This implies a stellar mass of about 1.7-2 M$_{\sun}$ from evolutionary tracks \citep{vda98}
that is compatible with the mass derived from CO observations. Our derived luminosity
of $\sim$ 8 L$_{\sun}$ implies a distance of $\sim$ 200-300 pc to reproduce the SED.
The Hubble Space Telescope images, as well as CO observations \citep{buj08}, suggest 
a disk radius of 4". Using the distance we derive of 200 pc this gives a physical radius of 800 AU,
consistent with current estimates of the outer radius for a disk around HAeBe objects (see compilation in 
\citet{hab04} and references therein). Furthermore, the A5-8 scenario does 
not necessitate the presence of a binary as suggested by \citet{buj08}.

Finally we note that the fit does not require the 8.2 $\mu$m BF commonly seen
in post-AGB stars (\citealt{job08}) and shows some PAH$^x$ emission. Thus, if not a PMS-star, GoHam would 
instead be a planetary nebula that can lack the 8.2 $\mu$m feature and show the
7.9 $\mu$m PAH$^x$ feature in their mid-IR spectrum.




\section{Summary and conclusion} \label{ext}

We have studied the mid-IR emission spectrum of 12 disks around young stars, showing clear signatures of stochastically heated
grains and no silicate emission. The observed spectra, emanating from very different objects, can be reproduced
using a mixture of only 5 template spectra, 3 of which are representative of interstellar (PDR) populations.
In addition, a PAH$^x$ population characteristic of highly UV-irradiated environments is present as is
a peculiar feature (dBF) possibly due to aliphatic material in analogy with the 8.3 $\mu$m broad feature
present in post-AGB stars. We can recapitulate the main
results of this work in three points:\\

(1) The mid-IR spectrum of disks traces the PDR created at the surface of the disk.
As in PDRs, we find that VSGs are destroyed to become free PAHs. VSGs and small PAHs survive long enough to be observed
only in the disks around cool stars. Around A stars only large PAHs have a lifetime that is long enough
to allow their detection. The coagulation of PAHs at the surface of disks, i.e. mid-IR emitting region, is likely hindered by 
the action of the UV field and therefore not correlated with the coagulation of larger grains. 
\\

(2) PAH$^x$ emission indicates the presence of a disk. This emission is found to be more prominant in HAe stars with high UV fields. 
This is compatible with PAH$^x$ being very large PAHs that can resist  photodestruction better. 
Spectroscopy indicates that they are charged and 
models favor cations rather than anions. The PAH$^x$ and dBF carriers are absent around Be stars, which we 
interpret as the absence of emission coming from a disk. This is likely because the disk is  small 
or event absent as it was photoevaporated.\\

(3) The shape of the mid-IR PAH spectrum of Herbig Ae/Be and T-Tauri stars is not related to disk morphology 
but to UV-irradiation conditions. This connection can be used to infer the spectral type / mass of the central source.
Using this result, we show that IRS 48 is of early A spectral type or produces a large amount of UV photons
in excess due to accretion. Similarly, we constrain the nature of the central source of GoHam 
and show that the mid-IR emission spectrum of this source is compatible with a protoplanetary 
disk with a central star of spectral type around A5-8.\\

Finally, we emphasize that detailed studies of a large sample of disks with PAH emission
are needed to provide deeper insights into the evolution of these very small dust particles along with the
evolution of the disk. Furthermore, spatially resolved spectro-imagery, in the aromatic 
emission range, of such objects will be crucial for the understanding of local evolution
of the disk, leading to planet formation.

\begin{acknowledgements}

This work was supported by the French National Program, Physique et Chimie du Milieu Interstellaireâ 
which is gratefully acknowledged. 

\end{acknowledgements}

\bibliographystyle{aa}
\bibliography{biblio}

\end{document}